\documentclass[12pt,letterpaper]{article} 
\usepackage{mrs2005,epsfig}
\usepackage{setspace}                     
\newcommand{\ie}{\textit{ie}}
\newcommand{\eg}{\textit{eg}}

\newcommand{\wwp}{\ensuremath{w_\mathrm{p}}}
\newcommand{\rrp}{\ensuremath{r_\mathrm{p}}}
\newcommand{\rrf}{\ensuremath{r_\mathrm{f}}}
\newcommand{\NNf}{\ensuremath{N_\mathrm{f}}}
\newcommand{\ttdyn}{\ensuremath{t_\mathrm{dyn}}}
\newcommand{\sigmav}{\ensuremath{\sigma_v}}
\newcommand{\GammaLRG}{\ensuremath{\Gamma_\mathrm{LRG}}}
\newcommand{\nnLRG}{\ensuremath{n_\mathrm{LRG}}}

\newlength{\figurewidth}
\begin{document}

\title{WHAT BEST CONSTRAINS GALAXY EVOLUTION IN THE LOCAL UNIVERSE?}

\author{DAVID W. HOGG}
\affil{Center~for~Cosmology~and~Particle~Physics,
       Department~of~Physics, New~York~University, 4~Washington~Pl.,
       New~York,~NY~10003,~USA.\\
       \texttt{david.hogg@nyu.edu}}

\begin{abstract}
After a polemical introduction about the proper activity of an
astrophysicist facing a dominant theoretical model and many Tb of
highly informative data, I review a few recent results on the
properties of galaxies in the nearby (redshift one-tenth) Universe
that directly bear on physical cosmology.  In one example, I show that
there are a number of ways of measuring, or strongly constraining,
massive galaxy--galaxy major merger rates, which are predicted with
limited uncertainties in the current generation of models.  In
another, I show that we can go beyond ``correlations'' between
individual galaxy properties and ``environment''.  Our results show
that it is galaxy star-formation histories---not their
morphologies---that are sensitive to environmental density.  I look
forward to a future, perhaps not that far away, in which these results
guide a fundamental modification to our theoretical assumptions,
though I fear that the dominant paradigm may not require subversion.
\end{abstract}

\section{Introduction}
\subsection{There's a lot of information out there}

\begin{figure}[htb]
\begin{center}
\epsfig{figure=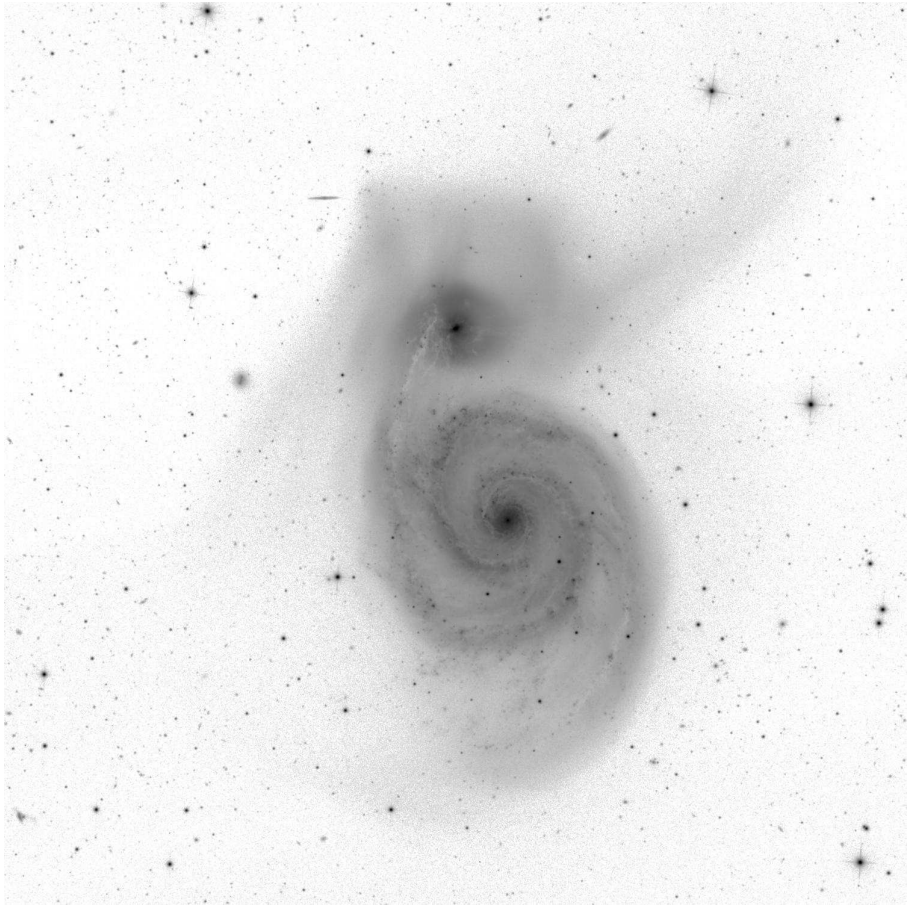,width=\figurewidth}
\end{center}
\caption{A black-and-white representation of SDSS visible imaging of
  M51.  Note the tidal tails, relationship of the spiral structure to
  the interaction, and the regularity of the dust.  This looks much
  better in color, of course.\label{fig:m51}}
\end{figure}

Figure~\ref{fig:m51} shows a (sorry, black-and-white) image of M51
taken from the Sloan Digital Sky Survey \cite{york00a} imaging data.
It positively \emph{teems} with \emph{information}.  You can see not
just a bulge and spiral arms, but (in the color version) that star
forming regions and dusty regions are systematically displaced within
the spiral arms, that the spiral galaxy is interacting with a smaller
bulge-dominated galaxy, that there are tidal tails and loops at very
low surface brightnesses, that the spiral structure is related to the
interaction, and even that the bulge-dominated galaxy contains an
active nucleus.  One of the ironies of my life is that these
fantastically beautiful, detailed, and informative pictures are almost
\emph{completely useless} for constraining physical models for the
formation and evolution of galaxies in a cosmological context.  We
simply do not have theoretical models---simulations---that produce
anything that looks even remotely like Figure~\ref{fig:m51}.  As you
read the contributions to this volume, many of which speak to the
great success of our cosmological model and the tremendous advances
made in the last ten years in understanding galaxy evolution, I would
like you to keep in mind this important point: \emph{These great
theories do not make M51.}  Of course, that is not to say they won't,
someday.

If that is one of the ironies of my life, then one of the
\emph{crusades} of my life is to promote an understanding of the
tremendous amount of astronomical information available to us, and the
best ways to harness that information in the service of physics.  By
``information'' here I don't just mean the touchy--feely thing you get
from reading this volume, but I mean the scalar quantity, measured in
bits, first worked out for the world by Shannon \cite{shannon49a} (in
what must be one of the most important developments in the entire
history of science).  I'll come back to this later, as it bears on
questions we will encounter, such as ``should I measure the age--mass
relation or the color--magnitude relation?'', and ``should I extract
the mean color as a function of environment, or the mean environment
as a function of color?''.

Philosophically, The primary goal of observational astrophysics---as
distinct from, say, pure astronomy---ought to be to \emph{rule out}
theoretical physical models.  Note that all important physical
experiments have the property that they ruled out---or could have
ruled out---one of the fundamental, dominant theories of their day.
In physical cosmology, we have a fundamental, dominant theory for the
growth of structure: Cold Dark Matter (CDM).  Our role as observers is
not to \emph{bolster} this model, or find ad-hoc parameters we can add
to the model that \emph{make it consistent} with the data.  Our role
is to find experiments that have the power, even in the face of
uncertainties (about, \eg, how galaxies form), to rule out or
substantially modify the fundamental assumptions of this theory.  If
an experiment does not have the power to rule out the theory, then it
can hardly be said to provide substantial support for it if its
results end up being in agreement!

That said, CDM may indeed turn out to be correct.  It is
\emph{certainly} correct on large scales, indeed with its successful
prediction of the angular spectrum of CMB inhomogeneities (yes, it was
truly a prediction, a parameterized prediction, and the results
\emph{do} live in that parameterized but nonetheless highly
constraining prediction space), the amplitude of the galaxy power
spectrum, and the baryon feature now detected at low redshift, CDM is
one of the best-tested theories in all of the physical sciences.  But
\emph{these tests are all on large scales} ($>10~\mathrm{Mpc}$).  On
scales the size of galaxies and galaxy clusters, CDM is \emph{not}
well tested because on these scales the star-formation,
black-hole-accretion, thermodynamic, and dissipative-evolution
uncertainties come in.  So lets harness the enormous power (the
information) of our observations, and the rapidly improving power of
our modeling, to make rock-hard tests of CDM on galaxy scales.

\subsection{Why work at a redshift of one-tenth?}

The principal disadvantage of working at low redshift is that the
redshift \emph{range} is small.  Although we \emph{do} see evolution
(in, \eg, the galaxy luminosity function, or mean galaxy specific
star-formation rates) \emph{within} low-redshift samples like SDSS
\cite{york00a} and 2MASS \cite{skrutskie97a}, we can measure evolution
directly with much greater precision by comparing samples at very
different redshifts.

On the other hand, the advantages of working at low redshift are many,
and they all relate to the enormous amount of information we have
about this epoch in the Universe.  With SDSS, 2MASS, and GALEX
\cite{martin05a} we have large enough solid angle and high enough
sensitivity to see a large fraction of the galaxy population in a
large fraction of the total volume of the Universe out to a redshift
of one-tenth.  In this review, I will focus on things we have learned
from the SDSS data, simply because that is what I know best.

The SDSS is enormously \emph{over-designed.}  The spectra in the SDSS
have far higher signal-to-noise than is necessary to obtain a
redshift; indeed we measure not just redshifts but star-formation
rates \cite{kauffmann03a, quintero04a}, stellar population mixtures
\cite{quintero04a}, and dust attenuations (both those affecting the
lines \cite{tremonti04a} and those affecting the stars
\cite{kauffmann03a}).

But that's not all!  The imaging in which these galaxies were selected
for spectroscopy is far deeper than necessary for object selection; a
typical Main Sample \cite{strauss02a} galaxy target is detected in the
SDSS imaging at a $S/N$ of many hundreds.  This means that we have
very high-precision galaxy colors and magnitudes, sizes,
concentrations, and surface brightnesses \cite{hogg02a, blanton03d}.

Finally, and importantly for what follows, the large contiguous area
of the SDSS (and other surveys) permits the study of clustering and
galaxy environments; it also allows us to identify large,
gravitationally bound systems, such as groups, clusters, and
superclusters.  Survey edges do not technically \emph{prevent} the
measurement of clustering and environments, but in practice, they make
it difficult.  So with the huge ``volume-to-boundary'' ratio of the
SDSS, we have good-quality measures of galaxy environments, the mean
galaxy density, and the correlation function on all scales out to the
scale of large-scale homogeneity \cite{hogg05a, eisenstein05b} (these
studies also demonstrate that the photometric calibration of the
survey is very stable on large angular scales).

\section{First example: galaxy--galaxy merger rate}

Recall our goal: \emph{Rule out the dominant theory.}  And recall our
approach: \emph{Use the information in the observations.}  What do
these mean in practice?  They mean that we should find the most robust
predictions in the dominant theory, and find the ``most informative''
observational test.  In principle, \emph{any} observation is a test of
theory, in that, in principle, it can be predicted.  However, some
predictions rely strongly on uncertain physics, including things like
star-formation triggering and efficiency, mechanical stellar feedback,
metallicity history, ionization history, and black-hole accretion
feedback.  Each of these processes affect the observed properties of
galaxies at a level the magnitude (and sometimes even the sign!) of
which is extremely uncertain.  The necessity of, in effect,
``marginalizing'' over these uncertainties removes almost all of the
information in most observational data (or, equivalently, removes
almost all of the precision of the prediction).  So we want to find
the observational tests that depend \emph{least} on the unknown
physical effects, and depend \emph{most} on the fundamental physical
model we want to attack.

In CDM, collapsed objects grow by accretion and merging.  This process
can be predicted with great precision for the dark sector, both
analytically and with numerical simulations.  Baryons only affect
these predictions on the smallest scales, smaller than the ``virial
radii'' of collapsed objects at the current day.  So as long as we can
find galaxies that are reliably associated with high-mass dark halos,
we can use the statistics of their merging to very directly test the
merging activity that CDM \emph{requires}.  The principal
uncertainties are related to the relationship between galaxies and the
halos that host them; this uncertainty is serious but \textsl{(a)}~it
is a limited, clearly circumscribed problem, and \textsl{(b)}~there
are many independent methods for investigating it, especially in the
many approaches of those working on the Halo Occupation Distribution
model for galaxy clustering \cite{scoccimarro01a, zehavi04a,
abazajian05a, berlind06a}.

\subsection{Close pairs}

If we know anything about the relationship between galaxies and
dark-matter halos, we know that the luminous red galaxies (LRGs)
\cite{eisenstein01a} in SDSS (super-$L^\ast$ red, dead, early-type
galaxies) live in very massive halos.  Indeed, their abundance and
clustering \cite{zehavi05a} very strongly constrain this relationship.
In addition, they contain very simple and very similar
(object-to-object) stellar populations \cite{eisenstein03b}, they have
uniform mass-to-light ratios, and they don't tend to have difficult
(for automated software) morphological features such as bars, HII
regions, dust lanes, or spiral arms.  LRGs are the natural galaxies to
use for the study of merger rates, because they are ``easy'' to
observe, and (relatively) easy to predict.

The only thing not ``easy'' about observing LRGs is that they are
rare, so you need an enormous volume.  The SDSS special LRG
spectroscopic sample \cite{eisenstein01a} fills $\sim
0.7~h^{-3}\,\mathrm{Gpc^3}$ and contains, even after conservative
cuts, $5\times 10^4$ LRGs \cite{hogg05a,eisenstein05b}.

\begin{figure}[htb]
\begin{center}
\epsfig{figure=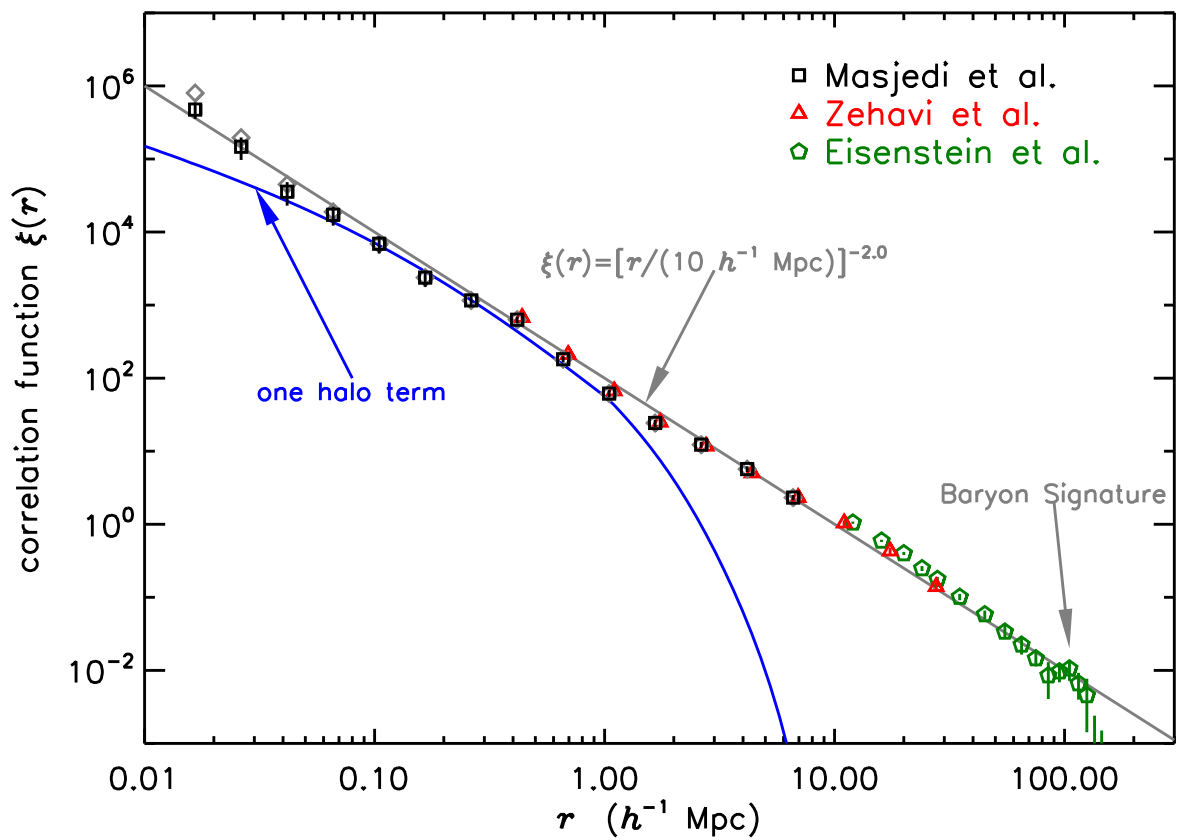,width=\textwidth}
\end{center}
\caption{Three-space correlation function for LRGs, deprojected from
  an estimate of the projected function $\wwp(\rrp)$ found by
  cross-correlating the spectroscopic and imaging samples, weighting
  to correct for the incompleteness due to SDSS spectrograph
  constraints or ``fiber collisions'' (from \cite{masjedi06a}).  Also
  shown are a hand-drawn power-law, a NFW profile \cite{navarro97a},
  and LRG correlation functions at intermediate scales
  \cite{zehavi05a} and large scales \cite{eisenstein05b}, the latter
  being a redshift-space function and therefore biased high on
  intermediate scales.\label{fig:xi}}
\end{figure}

We are looking for close pairs; the usual technical problem is that
there can be chance superpositions.  In principle we could use only
close pairs where both have SDSS redshifts that agree (within a
reasonable tolerance) but in practice we cannot (easily) because in
3/4 of the SDSS solid angle no two objects both get spectra if they
are closer than a hardware-enforced limit of 55~arcsec (the other 1/4
of the survey area is covered by ``overlap regions'' of the
spectroscopic fields and this repeat coverage essentially eliminates
this constraint).  We have excellent survey geometry and calibration,
though, so we don't merely count close pairs, we cross-correlate the
SDSS spectroscopic LRG sample with LRG targets in the imaging (whether
or not they got spectra), in the two-dimensional plane of the sky, but
scaled, for each spectroscopic LRG, to physical distances using the
redshift of that galaxy (and in cross-correlating, we statistically
remove chance interlopers at the ``subtract the mean density'' step).
We then spherically deproject this projected $\wwp(\rrp)$ to the
three-dimensional, real-space correlation function $\xi(r)$ shown in
Figure~\ref{fig:xi}.  Details are presented elsewhere \cite{masjedi06a}.

I find this result remarkable in that the correlation function is so
close to a power law over so many orders of magnitude, spanning the
range ($\sim100$~Mpc) where the structure formation is linear and
shows the baryon feature, through scales ($\sim 10$~Mpc) where
nonlinearities become important into the range ($\sim 1$~Mpc) at which
galaxy pairs tend to be within the same virialized, gravitationally
bound system, all the way down to galaxy-sized scales ($\sim 10$~kpc)
at which dynamical friction, tidal forces, and dissipation must
matter, at some level.  But those are subjects for an entirely
\emph{different} review.

Because of dynamical friction (among other things), sufficiently close
pairs of similar-mass LRGs (almost all of our pairs are similar in
mass, because the LRG luminosity selection is so strong) will merge in
a dynamical time.  In this picture, we can treat the correlation
function $\xi(r)$ at the smallest scales as a quasi--steady-state
inflow leading to the mergers of pairs of LRGs; we can
straightforwardly turn the measured $\xi(r)$ into a \emph{merger
rate}.  Really we put a \emph{limit} on the merger rate, because not
\emph{all} close pairs will merge in a dynamical time; they might take
much longer to merge, although they \emph{cannot} take much less.

There is some length scale $\rrf$ (the exact value is not important,
see below) inside of which dynamical friction is so effective that
pairs at this separation merge in a dynamical time $\ttdyn$, where the
dynamical time is
\begin{equation}
\ttdyn \approx \frac{2\pi\,\rrf}{\sigmav} \quad ,
\end{equation}
and $\sigma_v$ is some characteristic gravitational velocity for the
LRG in question.  The average number $\NNf$ of neighbor LRGs within
distance $\rrf$ of any ``target'' LRG is
\begin{equation}
\NNf \approx 4\pi\,\rrf^3\,\nnLRG\,\xi(\rrf) \quad ,
\end{equation}
where $\nnLRG$ is the space density of LRGs, and I have implicitly
used that $\xi(r)\sim r^{-2}$ and $\xi(\rrf)\gg 1$.  The LRG merger
rate $\GammaLRG$ is, therefore
\begin{equation}
\GammaLRG < \frac{\NNf}{\ttdyn}
   \approx 2\,\rrf^2\,\sigmav\,\nnLRG\,\xi(\rrf)
   \quad .
\end{equation}
Since Figure~\ref{fig:xi} shows something very close to $\xi(r)\sim
r^{-2}$, this expression for $\GammaLRG$ does not depend strongly on
the choice of $\rrf$, which is indeed poorly known.

This merger rate (LRG mergers per LRG) can be written as
\begin{equation}
\GammaLRG < \frac{1}{160~\mathrm{Gyr}}
   \,\left[\frac{\rrf^2\,\xi(\rrf)}{(10~\mathrm{kpc})^2\,10^6}\right]
   \,\left[\frac{\sigmav}{300~\mathrm{km\,s^{-1}}}\right]
   \,\left[\frac{\nnLRG}{10^{-4}~\mathrm{Mpc^{-3}}}\right]
   \quad ,
\end{equation}
\ie, each LRG has less than a one percent probability of merging with
another LRG each Gyr.

A few points deserve emphasis: \textsl{(a)}~Unlike most of the results
in the literature, there are no worries here about chance
superpositions contaminating the results; we have deprojected to the
three-space $\xi(r)$.  We have the volume and the numbers to make that
deprojection believeable.  \textsl{(b)}~This is a correlation
function, not a ``pair fraction,'' so the number density has been
divided out.  The pair fractions commonly discussed in the literature
have the ``instability'' that they are proportional to the abundance
(of secondaries) and that abundance can be an extremely strong
function of magnitude, color, and surface-brightness cuts.  Indeed,
some of the discrepancies in the literature about pair fraction and
its evolution may be caused by these instabilities.  \textsl{(c)}~This
rate is a \emph{strict upper limit}; two galaxies in a pair simply
cannot in a time much shorter than the local dynamical time.

Our LRG--LRG rate limit ($< 1$~percent per Gyr) is low, much lower
than the rate at which LRG-hosting dark-matter halos merge with one
another.  On the other hand, the galaxies \emph{in} halos merge much
less frequently than the halos themselves.  One frustrating aspect of
the literature is that although there are some theoretical predictions
\cite{murali02a}, none of them is presented in quite the way you need
to do a direct comparison.  But we are very optimistic that this rate
can be directly compared to models, for all the reasons given above
(LRGs are simple and massive and in massive halos).  We are also
optimistic that we can repeat this experiment for a wide range of
galaxy masses and mass ratios.

\subsection{Post-starburst galaxies}

Any measurement of a merger rate involves a measurement of an
\emph{abundance} of candidates and a \emph{timescale} on which they
will merge, are merging, or have merged.  Often the timescale is quite
uncertain, but there is one place where it is well known: in the
passive evolution of starbursts.

\begin{figure}[htb]
\begin{center}
\epsfig{figure=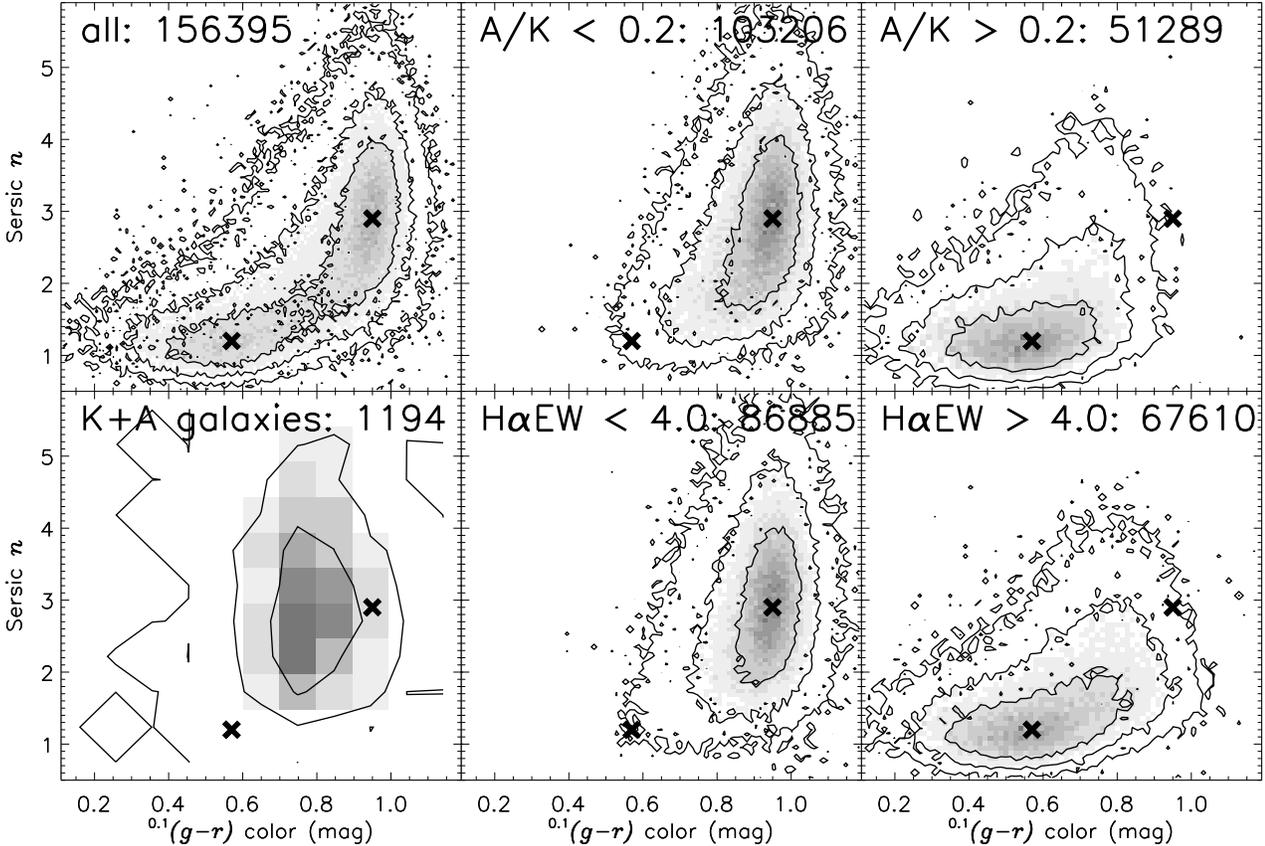,width=\textwidth}
\end{center}
\caption{The distribution of all galaxies, star-forming galaxies, dead
  galaxies, and post-starburst (K+A) galaxies in the space of color
  and concentration (from \cite{quintero04a}).  The top left panel
  shows all galaxies, with crosses superimposed to guide the eye to
  the early-type (red, concentrated) and late-type (blue, exponential)
  galaxy populations, which separate nicely in this plane.  The four
  panels on the right show the same for various cuts on A-star-based
  and OB-star-based measures of the recent-past star-formation
  history.  The lower left panel shows the same for the post-starburst
  population.  Note how they are concentrated and blue, as they will
  ``fade into'' the early-type location.\label{fig:k+a}}
\end{figure}

There are two independent arguments---beyond the simple argument that
stars probably form in disks and mergers are required to destroy those
disks---that post-starburst galaxies lie along an ``evolutionary
sequence'' connecting disk-dominated galaxies (spirals and late-type
galaxies and irregulars) to bulge-dominated galaxies (lenticulars and
ellipticals) via merging (that triggers the burst and disrupts the
disks).  The first is that bulge-dominated chemical abundance patterns
differ from late-type patterns, and the difference can be explained if
star-formation terminates with a large, rapid burst that exhausts all
remaining fuel for star formation \cite{worthey98a}.  The second is
that the photometric properties of post-starburst galaxies are such
that their passive fading will cause them to rapidly evolve \emph{not}
to the properties of disks, but to the properties of bulges.  We
demonstrated this latter point with a sample of post-starburst
galaxies taken from the SDSS \cite{quintero04a} (see
Figure~\ref{fig:k+a}).

By making quasi-steady-state models of the distribution of A-star
excesses in these galaxies, we can determine a \emph{rate} density of
their formation; we get $\sim
10^{-4}\,h^3~\mathrm{Mpc^{-3}\,Gyr^{-1}}$, or a bulge-dominated galaxy
population that is growing by $\sim 1$~percent per Gyr.

What is this the rate \emph{of}?  Since some mergers are no-doubt
``dry'' (such as the LRG--LRG mergers discussed above), we can see
this as a lower limit to the merger rate, or the bulge-dominated
galaxy population growth rate.  However, since some of these
starbursts might be triggered by very small accretion events or even
tidal or internal perturbations, we can \emph{also} see this as an
upper limit.  With so many uncertainties, why do I bring this up in a
review of \emph{robust} tests of fundamental physics?  Because the
data are highly \emph{informative,} providing a relatively precise
rate density.  Our work now is concentrated on constraining the nature
of the triggering events; if we can work this out we might have a very
strong constraint on merger activity in the local Universe.

\subsection{Tidal features}

The most straightforward method for measuring the merger rate is to
simply look through an atlas of galaxies and identify galaxies that
are clearly undergoing, or recovering from, a merger.  There are many
signatures, of varying veracity, including tidal arms, ``S''-shaped
distortions, shells, bars, warps, boxiness, and minor-axis dust lanes.
I am very sympathetic to this procedure, and it pains me that it has
not been executed definitively in the huge datasets we now have available
to us in the nearby Universe.

That said, the morphological method, while straightforward, does not
necessarily give you the best measure of the merger rate.  Recall
again that we need an \emph{abundance} and a \emph{timescale.}
Identifying morphological signatures, and deciding on clear boundaries
between ``merger candidate'' and ``not'' is more of an art than a
science at this stage.  In addition, the timescales over which these
features are visible are not known, and they depend on the types of
features, the events that raised them, and the fraction of bolometric
output that comes from young stellar populations or dust-enshrouded
star-formation.  That said, there has been some remarkable
quantitative work on the ages of particular morphological features,
and reconstruction of individual galaxy--galaxy merger events
\cite{kaufman99a, salo00a, struck03a}.  There have also been some
attempts to make automated classification \cite{conselice03a}.

My problems with the ``morphological method'' for determing the merger
rate are many: \textsl{(a)}~Features of importance can have a wide
range of surface brightnesses; some appear in the UV (\eg, the
Antennae) and some appear in the near-IR (\eg, the Whirlpool in
Figure~\ref{fig:m51}).  \textsl{(b)}~If you look hard enough,
everything is ``disturbed.''  Indeed it has been discussed at this
meeting that at ``faint enough'' surface brightnesses, essentially
\emph{all} early-type galaxies show tidal features
\cite{vandokkum05a}.  Since (above) I constrained the major-merger
rate to be low, either these features are raised by very small (and
more numerous) accretion events or else they last an extremely long
time.  \textsl{(c)}~Not all asymmetries \cite{conselice03a} or
``shell-like'' features \cite{malin83a} are clear evidence for
merging, as both can be created by tidal perturbations, small
accretion events, localized star-formation episodes, and possibly
cooling flows from halo plasma.  \textsl{(d)}~Even the features that
\emph{are} clearly signs of mergers can last---observably---for
anything from the lifetime of an O star (a small fraction of a
dynamical time) to many dynamical times, depending on star-formation
activity and the ``coldness'' of the progenitors.

Fundamentally, until these issues are cleared up, I am bearish on
constraining CDM with morphological studies.

\section{Second example: galaxy environments}

Different parts of the Universe, with different local densities,
evolve with different ratios of gravitational-collapse,
star-formation, and galaxy-dynamical timescales.

\subsection{Environment measures}

Unfortunately, there are no \emph{good} measures of local density!
Counting galaxies in close volumes is subject to small-number Poisson
statistics (or worse).  In principle, tesselations or transverse
distances to $N$th-nearest neighbor give precise numbers.  However,
these kinds of measures have ill-defined length-scale and are subject
to large jumps when sample definition (\eg, magnitude cut) is changed.

We have shown that all the information about environment we see is
captured by environment measures on $\sim 1$~Mpc scales
\cite{blanton04c}, which happens (not coincidentally, I imagine) to be
on the order of the virial scale.  This has recently been
controversial \cite{balogh04a}, but I think the point is settled now.

\subsection{Star-formation history and environment}

\begin{figure}[htb]
\begin{center}
\epsfig{figure=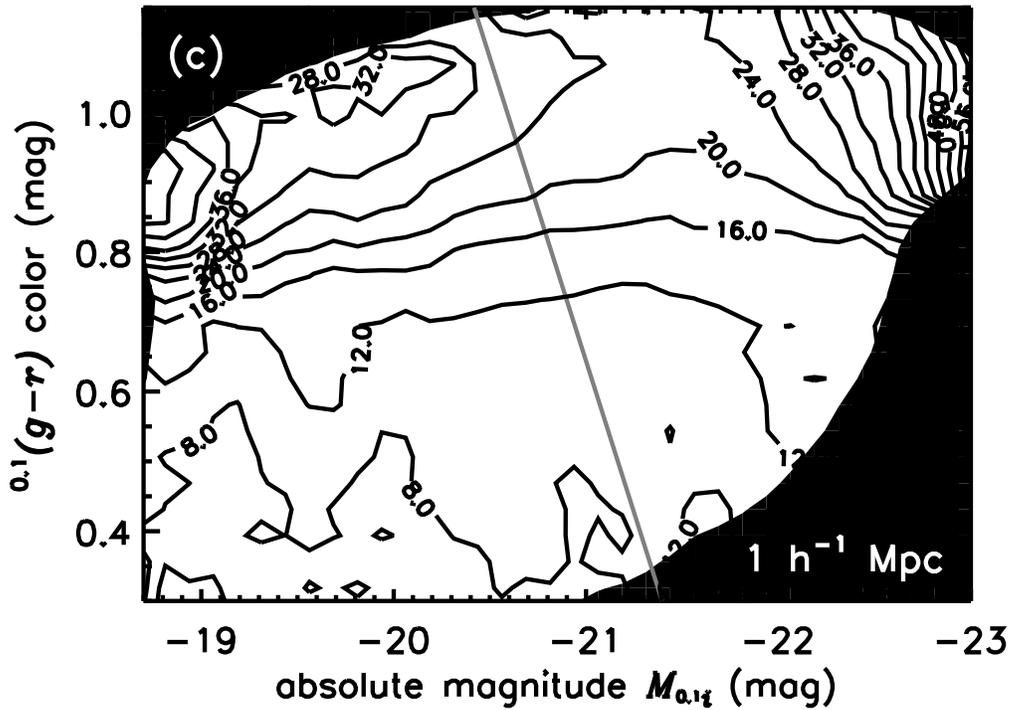,width=0.75\textwidth}
\end{center}
\caption{The mean overdensity (measured in a gaussian \emph{spherical}
  de-projected window \cite{eisenstein03a} of radius 1~Mpc) of the
  environments of galaxies as a function of galaxy color and
  luminosity (from \cite{hogg03b}).  The fact that the contours are
  close to horizontal over most of the plot shows that it is color,
  not luminosity, that is most closely related to environmental
  density for most of the luminosity range; luminosity only
  ``matters'' at high luminosity.\label{fig:cme}}
\end{figure}

It has been observed for some time now that specific star-formation
rates are different, on average, in different environments
\cite{kennicutt83a, hashimoto98a, balogh01a, martinez02a, lewis02a,
gomez03a, blanton03d}.  In CDM, the most \emph{massive} halos are in
the most dense environments, so shouldn't we expect \emph{total
stellar mass} to be an \emph{even stronger} function of environment?
As we emphasize here and below, we don't have to think about these
questions separately---or theoretically:

Figure~\ref{fig:cme} shows the mean environment as a function of color
and absolute magnitude.  It separates the current star-formation rate
(color) and the total time-averaged star-formation rate (luminosity),
and shows that it is the former, not the latter, that relates most
closely to environment around $L^{\ast}$.  Only the very most massive
galaxies show a strong relationship between total stellar mass and
environment.

In current models, the relationship shown in Figure~\ref{fig:cme} is
simply explained in terms of satellite and central galaxies in groups
\cite{berlind05a}; at most luminosities, most galaxies are central,
not satellites, but at fixed luminosity, the satellites tend to be
redder and in higher density environments (because at early times they
fell into halos massive enough to contain more than one galaxy of
their luminosity).

You might ask, in Figure~\ref{fig:cme}, why we averaged the
\emph{environment} as a function of galaxy properties, and not the
other way around.  I noted that environmental density can never be
measured at high signal-to-noise.  Color and magnitude, on the other
hand, are measured in SDSS data with percent precision (though perhaps
not percent \emph{accuracy}).  It would be insane to average away that
precision, and split the sample by a noisy environmental measure
(thereby effectively smoothing any important relationships by the
uncertainties in the environmental measures).  Most investigators feel
that it is somewhat more ``natural'' to look at galaxy properties as a
function of environment (\ie, the wrong way around), but my point is
that my way is \emph{more informative} (objectively), and any
theoretical model can predict the relationship either way!

You might also ask, in Figure~\ref{fig:cme}, why we showed the color
and luminosity dependences of environmental density, and not the
``age'' and ``stellar mass'' dependencies.  This issue is somewhat
subtle, and it depends on your goals, but I note the following: The
color-magnitude diagram of galaxies contains a lot of Shannon
\emph{information} in its narrow red sequence and less-narrow blue
sequence \cite{blanton03d}.  Any honest transformation to age and mass
involves modeling uncertainties that will (and do \cite{brinchmann04a,
padmanabhan04a}) ``smooth'' the diagram, destroying information.
Obviously, comparisons with theoretical models involve making these
uncertain transformations.  The question is, who should make them, the
theorist or the observer?  Personally, I respect the information in
the data; it is anathema to transform and lose this richness.

\subsection{There is no morphology--density relation (we can find)}

In some sense, the most important ``fact'' of galaxy evolution is the
observation that \textsl{(a)}~the morphological mix is different in
different environments \cite{dressler80a,postman84a}.  At the same
time, \textsl{(b)}~galaxy morphology is related to star-formation
history (and a lot of other stuff) \cite{tammann79a, kormendy89a,
roberts94a}.  I just showed that \textsl{(c)}~star-formation history
is related to environmental density.  Are all three relations on an
\emph{equal footing}, or is one the product of the other two?  We find
ourselves in the position of epidemiologists, trying to understand the
causes of asthma in the face of strong relationships between race,
socio-economic status, diet, and living conditions.  Fortunately we
have great data, and a lot of it, more and better than the
epidemiologists (I am sad to say).

\begin{figure}[htb]
\begin{center}
\epsfig{figure=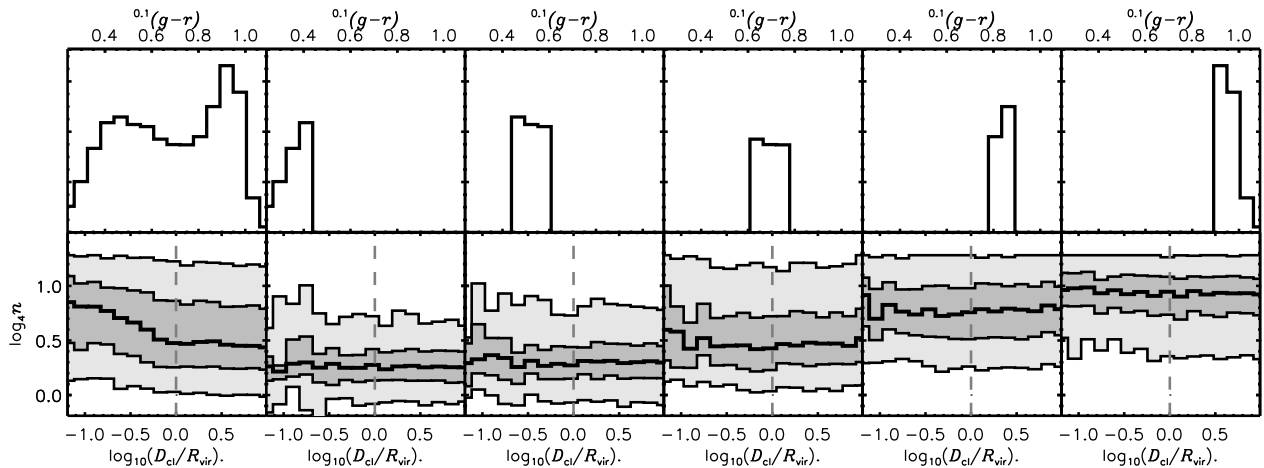,width=\textwidth}
\end{center}
\caption{The top left panel shows the color distribution for a
  subsample of the SDSS Main Sample galaxies with redshifts
  $0.015<z<0.068$.  The bottom left panel shows the dependence of
  weighted quantiles of concentration (S\'ersic index) on environment
  (clustocentric distance; details in \cite{quintero06a}).  The
  quantiles shown are 5~percent, 25~percent (first quartile),
  50~percent (median), 75~percent (third quartile), and 95~percent.
  Note the strong dependence of concentration on evironment in the
  bottom left panel.  In each of the next pairs of panels, the top
  shows the color distribution for a color-selected 20-percent
  quantile of the data, and the bottom shows the dependence of
  concentration on evironment within that color quantile.  Note that
  there is \emph{very little dependence of concentration on
  environment at a given color.}  (Figure adapted from
  \cite{quintero06a}.)\label{fig:ndontmatter}}
\end{figure}

\begin{figure}[htb]
\begin{center}
\epsfig{figure=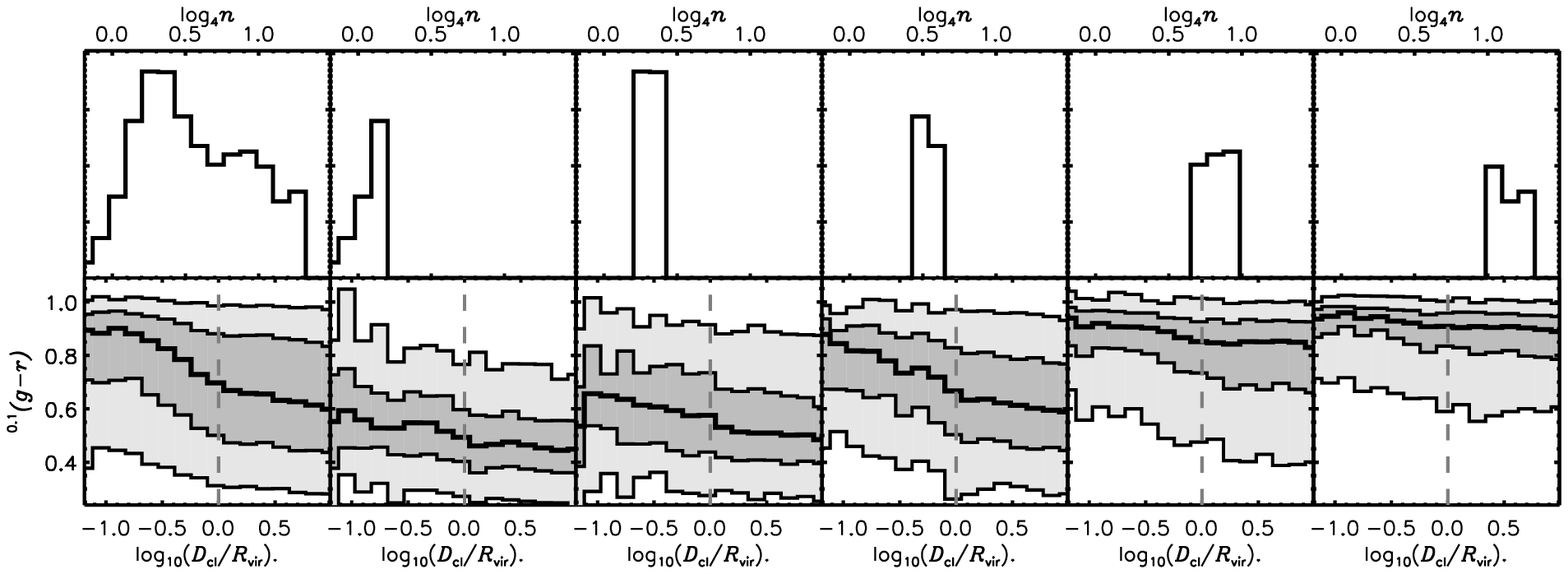,width=\textwidth}
\end{center}
\caption{The same as Figure~\ref{fig:ndontmatter} but with color and
  concentration reversed; \ie, this figure shows the dependence of
  color on environment within concentration-selected quantiles.  It
  shows that color \emph{does} depend on evironment, even at fixed
  concentration (adapted from
  \cite{quintero06a}).\label{fig:colormatter}}
\end{figure}

In Figure~\ref{fig:ndontmatter} I show the variation of concentration
(a measure of bulge/total ratio and therefore a morphology surrogate
\cite{strateva01a, goto03b}) on environment (clustocentric distance, a
high-precision but variable-scale environment measure), in narrow
color slices.  In Figure~\ref{fig:colormatter} I show the variation of
color on environment in narrow concentration slices.  These Figures
look very different: Color depends on environment independently of
morphology, morphology \emph{does not} independently of color.  Of
course I am over-stating the result by calling this ``morphology'';
take it as you wish, but it is clear that color and concentration are
\emph{not} on an equal footing when we ask what they can tell us about
environment.

In current thinking about the \emph{reasons} for the
morphology--density relation (things like ram-pressure stripping,
mergers, late accretion, tidal perturbations), these results are very
difficult to understand.  What physical processes can ``tell'' galaxy
star-formation rates about their environments and ``tell''
morphologies to keep track of star-formation rates but not do much to
the morphologies independently?  I think the conclusion has to be that
the processes that set the morphology (or, really, concentraion or
bulge/total ratio) are somehow \emph{internal} to the galaxies.
That's going to be hard to reconcile with CDM cosmogony.

\section{Future challenges}

\subsection{Short-term}

For the theorists, I have two requests: \textsl{(a)}~Explain or
predict our LRG--LRG merger rate and more merger rates to come.
\textsl{(b)}~I didn't discuss this, but convincingly
explain the luminosity function, especially its remarkable
difference from the mass function of collapsed objects.  After years
of work, this has still not been done \cite{benson03a}.

For the observers, I have two more: \textsl{(c)}~Separate the
kinematically cold and kinematically hot parts of galaxies, without
relying on the almost totally unjustified radial profiles used in
bulge/disk decomposition.  God didn't tell us that disks are
exponential, and he \emph{certainly} didn't tell us that bulges are
de~Vaucouleurs.  Your efforts will be rewarded, because the most
sophisticated modern theoretical models \emph{do} distinguish these
components; they make non-trivial predictions. \textsl{(d)}~Respond to
my objections (above) and measure the merger rate morphologically.

\subsection{Long-term}

Long term, we need to have simulations that \emph{look like galaxies}
on all the scales and at all the wavelengths we can observe.  Recall
Figure~\ref{fig:m51}.  Perhaps when these simulations exist we will
understand what sets galaxy morphologies.

If I am to set myself a \emph{very} long-term goal so absurdly
difficult it will take much longer than my own natural lifetime, a
goal that will make use of literally \emph{all} of the information in
the observations, and a goal that is the extrapolation to $t=\infty$
of the research program I have started here, it would be this:
Construct a chain of \emph{constrained realizations} of the entire
observable Universe, a chain (as in a Markov chain or equivalent) that
can be used to marginalize over \emph{all} unknowns, both in the data
and in the theory.  Each ``link'' in the chain would be a simulation
or model of the entire observable Universe, consistent with every
existing observation (\ie, it would get all the \emph{phases} right,
not just the amplitudes and dynamics).

It sounds ridiculous, but with the next (or next to next) generation
of multi-scale simulations, it might be possible.  Of course there
would be huge degeneracies, both because our observations are
incomplete and because large parts of the physical model are uncertain
(especially those parts relating to small-scale baryonic physics).
Each new observation or improvement in theoretical understanding would
rule out parts of the chain, and permit finer sampling in the (now
smaller) allowed regions of parameter space.  This chain of
realizations could be used to tell us the initial conditions,
evolutionary history (and a ``neighborhood'' around those still
consistent with the data and theory) for every \emph{actual} galaxy we
have observed, and provide a clear guide for further observational
testing.

Indeed, those predicted observables---things to observe but not yet
observed---that show enormous variation among the realizations in the
chain will be the decisive observations for the future.  Those
predicted observables that show the least variation will be the
\emph{most fundamental} predictions of the physical cosmology.  A
chain of constrained realizations is a tool for the \emph{objective}
pursuit of the objectives I set out in the introduction; it could
provide us with a quantitative ranking of the importance of each
possible observational experiment, a ranking by its relative
capability of \emph{ruling out the dominant physical theory.}  If we
can't rule it out, then we are just ``constructing'' it ``socially''.

\acknowledgements{All of this work was done in very close
  collaboration with Mike Blanton at NYU, to whom I owe enormous
  thanks; in addition, Andreas Berlind, Richard Cool, Daniel
  Eisenstein, Morad Masjedi, Jim Peebles, Alex Quintero, Sam Roweis,
  and Beth Willman contributed to various parts.  This research was
  supported by NASA (NAG5-11669) and the NSF (AST-0428465), made use
  of the NASA Astrophysics Data System, and employed open-source
  software written and maintained by David Schlegel, Doug Finkbeiner,
  Mike Blanton, Nikhil Padmanabhan, myself, and others.  All the data
  shown in this review have been released publicly by Mike Blanton and
  myself as part of the NYU-VAGC \cite{blanton05a}.}

\bibliographystyle{plain}
\bibliography{apj-jour,ccpp}

\vfill
\end{document}